\documentclass[aps,prl,twocolumn,showpacs,preprintnumbers,amsmath,amssymb]{revtex4-1}

\def\O{{\cal O} }

\def\N{{\cal N} }

\def\L{{\cal L} }
\def\J{{\cal J} }

\def\M{{\cal M} }

\def\dcut{{%
    \setbox0\hbox{D}%
    \rlap{\hbox to \wd0{\hss ~/ \hss}}\box0}}

\usepackage{graphicx}
\usepackage{dcolumn}
\usepackage{bm}
\usepackage{dsfont}
\usepackage{amssymb,amsmath}
\usepackage{epsfig}
\usepackage{epstopdf}
\usepackage{latexsym}
\usepackage{graphicx}
\usepackage{subfigure}
\usepackage{booktabs}
\usepackage{color}

\def\del{\nabla}

\def\p{\partial}

\newcommand{\be}{\begin{equation}}
\newcommand{\ee}{\end{equation}}
\newcommand{\bal}{\begin{align}}
\newcommand{\eal}{\end{align}}

\begin{document}

\title{Hall Scrambling on Black Hole Horizons}

\author{Willy Fischler$^{a,b}$, Sandipan Kundu$^{a,b,c}$}

\affiliation{$^{a}$Theory Group, Department of Physics, University of Texas, Austin, TX 78712}

\affiliation{$^b$Texas Cosmology Center, University of Texas, Austin, TX 78712}

\affiliation{$^c$Department of Physics, Cornell University, Ithaca, New York, 14853, USA}
\affiliation{E-mail: fischler@physics.utexas.edu, kundu@cornell.edu}

\begin{flushright}
\small{~~~~\\ ~~~~~~~~~~~~~~~~~~~~~~~~~~~~~~~~~~~~~~~~~~~~~~~~~~~~~~~~~~~~~~~ UTTG-31-14 TCC-031-14}\
\end{flushright}

\begin{abstract}
We explore the effect of the electrodynamics $\theta$-angle on the macroscopic properties of black hole horizons. Using only classical  Einstein-Maxwell-Chern-Simons theory in $(3+1)-$dimensions, in the form of the membrane paradigm, we show that in the presence of the $\theta$-term, a black hole horizon behaves as a Hall conductor, for an observer hovering outside. We study how localized perturbations created on the stretched horizon scramble on  the horizon by dropping a charged particle. We show that the $\theta$-angle affects the way perturbations scramble on the horizon, in particular, it introduces vortices without changing the scrambling time. This Hall scrambling of information is also expected to occur on cosmological horizons.

\end{abstract}

\maketitle

\section{I. Introduction and a summary}

Black hole horizons have eluded physicists ever since astronomer Karl Schwarzschild first found them in 1915 as a simple yet puzzling consequence of the general theory of relativity. The description of the near horizon region of a black hole in terms of quantum field theory  in curved spacetimes \cite{bd,Wald:1995yp} historically has provided us with a deep insight into the mysterious innards of quantum gravity, \emph{e.g.} thermodynamic descriptions of black holes. It was believed that quantum field theory in curved spacetimes provided a good physical description when quantum gravitational effects do not play a significant role, however this idea has recently been challenged \cite{fischler}.  Various puzzles and paradoxes \cite{Almheiri:2012rt,braunstein} indicate a possible failure of  quantum effective field theory in the near horizon region of a black hole.

String theory, Matrix Theory \cite{Banks:1996vh}, and the AdS/CFT correspondence \cite{Maldacena:1997re}, which are the only models of quantum gravity over which we have mathematical control, strongly suggest that black hole evolution as seen by an external observer, obeys the usual rules of unitary quantum mechanics, and verifies the Bekenstein-Hawking entropy formula in a large class of special cases.  Thus far none of them give us a comprehensive description of the physics of the black hole, and in particular, there is no definitive clue about the microscopic description of the near horizon region of a black hole. 

On the other hand, the {\it membrane paradigm} \cite{Price:1986yy,Thorne:1986iy} provides a powerful framework to study macroscopic properties of black hole horizons by replacing the true mathematical horizon by a {\it stretched horizon}, an effective time-like membrane located roughly one Planck length away from the true horizon. Indeed, in astrophysics the membrane paradigm has been successfully used to study phenomena in the vicinity of black holes, outside their horizons (see \cite{Thorne:1986iy} and references therein). Macroscopic properties of black hole horizons can also provide crucial hints about details of the microscopic physics \cite{Sekino:2008he}. Predictions of the membrane paradigm are generally considered to be robust since they depend on some very general assumptions: (i) the effective number of degrees of freedom between the actual black hole horizon and the stretched horizon are vanishingly small, (ii) physics outside the black hole, classically must not be affected by the dynamics inside the black hole.

Quantum chromodynamics (QCD) is an integral cornerstone of our understanding of particle physics. It is well known that Lorentz and gauge invariance allow QCD action to have a CP-violating, topological $\theta_{QCD}$ sector \cite{Crewther:1979pi}
\be
\L_{\theta}=\frac{\theta_{QCD}g^2}{64 \pi^2}\epsilon^{\alpha \beta \mu \nu}F^a_{\alpha \beta}F^a_{\mu \nu}\ ,\nonumber
\ee
where $F^a_{\alpha \beta}$ is the field strength and a priori parameter $\theta_{QCD}$ can take any value between $-\pi$ and $\pi$. This topological term can have physical effects, for example  it contributes to the neutron electric dipole moment. Experiments set a rather strong limit: $|\theta_{QCD}|<<10^{-9}$, which indicates that this term is unnaturally small.  Similarly, Lorentz and gauge invariance also allow the electrodynamics action to have a CP-violating $\theta$ term
\begin{align}
S=\int d^4x \left[- \frac{\sqrt{g}}{4} F_{\mu\nu}F^{\mu \nu} +\frac{\theta}{8}  \epsilon^{\alpha \beta \mu \nu}F_{\alpha \beta}F_{\mu \nu}\right]\ .\nonumber
\end{align}
The electrodynamics $\theta$-term is an elusive quantity; it does not affect the classical equations of motion because it is a total derivative. Therefore, it does not contribute for perturbative quantum electrodynamics (QED), which strongly indicates that the effects of the $\theta$ term in QED, if any, are non-perturbative, making it difficult to detect. However, as Witten showed \cite{Witten:1979ey}, in the presence of illusive magnetic monopoles, this term can have a measurable physical effect because it provides monopoles with electric charges proportional to $\theta$.   In this paper, by coupling this theory to gravity  we will show that the $\theta$-term can affect the electrical properties of the black hole horizon.  In particular, in the context of the membrane paradigm by using only classical { \it Einstein-Maxwell-Chern-Simons theory} in $(3+1)-$dimensions,  we argue that in the presence of the $\theta$-term, a black hole horizon behaves as a {\it Hall conductor} with Hall conductance $\approx \theta(377\Omega)^{-1}$.

We will provide, in this paper, evidence for this behavior by analyzing a simple {\it thought experiment}, in which an outside observer drops a charge onto the black hole and watches how the perturbation scrambles on the black hole horizon. In quantum mechanics, information contained  inside a small subsystem of a bigger system, is said to be scrambled when the small subsystem becomes entangled with the rest of the system and the information can only be recovered by examining nearly all the degrees of freedom of the system. It is indeed remarkable, as pointed out by Sekino and Susskind, that black hole horizons (and also de Sitter horizons) are the {\it fastest scramblers} in nature \cite{Sekino:2008he,Susskind:2011ap}; in particular, for a black hole with temperature $T$ and entropy $S$, the scrambling time goes as $t_s T \approx \hbar \ln S$. This ``{\it fast-scrambling}" on black hole and de Sitter horizons strongly indicates that the microscopic description of scrambling of information on static horizons must involve non-local degrees of freedom \cite{Sekino:2008he,Susskind:2011ap}. We will show that the $\theta$-angle affects the way charges scramble on the horizon, in particular, it introduces vortices without changing the scrambling rate -- we will call this phenomenon ``{\it Hall scrambling}". Since Hall scrambling depends only on the {\it Rindler-like} character of the black hole horizon but not on the details of the metric, the same conclusion is also true for arbitrary cosmological horizons. A microscopic description of fast-scrambling should be able to explain the origin of Hall-scrambling in the presence of the $\theta$-angle.

It is very tempting to apply the framework of membrane paradigm in the context of {\it holography}. In fact the {\it AdS/CFT correspondence} \cite{Maldacena:1997re} has taught us that the low frequency limit of linear response of a strongly coupled quantum field theory is related to that of the membrane paradigm fluid on the black hole horizon of the dual gravity theory \cite{Kovtun:2003wp,Son:2007vk,Iqbal:2008by}. This raises a deeper question. Can we then conclude that the same has to be true for holographic models of cosmological spacetimes?
 
The rest of the paper is organized as follows. We start with a brief discussion of the membrane paradigm in section II. In section III, we show that in the presence of electrodynamics $\theta$-term, the black hole horizon behaves as a Hall conductor. Then in section IV, we introduce Hall scrambling for Rindler, black hole and cosmological horizons. Finally, we make some comments in section V about its connection with the AdS/CFT correspondence and  conclude in section VI.  

\section{II. Membrane paradigm}
Let us begin with a brief discussion of the black hole membrane paradigm \cite{Price:1986yy,Thorne:1986iy}. Finiteness of the black hole entropy indicates that at a distance less than the Planck length from the black hole horizon, the effective number of degrees of freedom should be vanishingly small. So, it is more natural as well as more convenient to replace the true mathematical horizon by a stretched horizon, an effective time-like membrane $\M$ located roughly one Planck length away from the true horizon.

Advantages of having a stretched horizon will be more apparent if we consider some fields in the black hole background with an action 
\begin{equation}\label{action}
S_{tot}=\int d^{d+1}x \sqrt{g}\L(\phi_I, \del_\mu \phi_I)\ ,
\end{equation}
where $\phi_I$ with $I=1,2,...$ stands for any fields. It is necessary to impose some boundary conditions on the fields $\phi_I$ in order to obtain equations of motion by varying this action. We will impose  Dirichlet boundary conditions $\delta \phi_I =0$ at the boundary of space-time. The stretched horizon $\M$ divides the whole space-time in regions: $A$ (outside the membrane $\M$) and $B$ (inside the membrane $\M$). Therefore, we can write $S_{tot}=S_A+S_B$. 

Now imagine an observer $\O$ who is hovering outside the horizon of a black hole. Observer $\O$ has access only to the region outside the black hole and physics he observes, classically must not be affected by the dynamics inside the black hole. That means observer $\O$ should be able to obtain the correct equations of motion by varying some action $S_\O$ which is restricted only to the space-time outside the black hole. Clearly, $S_\O \neq S_A$ because the boundary terms generated on $\M$ from $S_A$ are in general non-vanishing. Therefore, we should add some surface terms that exactly cancel all these boundary terms. Let us now rewrite the total action in the following way \cite{Parikh:1997ma}
\be
S_{tot}=\left(S_A + S_{surf} \right)+ \left(S_B - S_{surf} \right)\ ,
\ee
such that $S_A + S_{surf}$ and $S_B - S_{surf}$ are independent of each other and correct equations of motion can be obtained by varying them individually. For the observer $\O$, the action  $S_\O=S_A + S_{surf}$ for fields $\phi_I$  now have sources residing on the stretched horizon
\begin{align}
S_\O=&\int_A d^{d+1}x \sqrt{-g}\L(\phi_I, \del_\mu \phi_I)\nonumber\\
 &~~~~~~~~~~~~+\sum_I \int_\M d^{d}x \sqrt{-h} \ \J_\M^I \phi_I
\end{align}
where, $h$ is the determinant of the induced metric on the stretched horizon $\M$ and sources are
\be\label{source}
\J_\M^I= \left[n_\mu \frac{\partial \L}{\partial \left(\del_\mu \phi_I \right)}\right]_\M
\ee
where, $n_\mu$ is the outward pointing normal vector to the time-like stretched horizon $\M$ with $n_\mu n^\mu =1$. The observer $\O$ who is hovering outside the horizon, can actually perform real experiments on the stretched horizon $\M$ to measure the sources $\J_\M^I$. It is important to note that one should interpret $\J_\M^I$ as external sources such that $\frac{\delta \J_\M^I}{\delta \phi_J}=0$.

\subsection{A. Electromagnetic fields and stretched horizon}
The action for electromagnetic fields in the curved space-time in $(3+1)-$dimensions is
\be\label{emaction}
S=\int \sqrt{g}d^4x \left[- \frac{1}{4}F_{\mu\nu}F^{\mu \nu}+j_\mu A^\mu \right]\ ,
\ee
where, as usual $F_{\mu\nu}=\p_\mu A_\nu-\p_\nu A_\mu $ and $g=-\det\left(g_{\mu\nu}\right)$.  $j^\mu$ is a conserved current, i.e., $\del_\mu j^\mu=0$. Our convention of the metric is that the Minkowski metric has signature $(-+++)$. The equation of motion obtained from action (\ref{emaction}) is
\begin{align}\label{meq}
\del_\mu F^{\mu\nu}=-j^\nu\ .
\end{align}
Field strength tensor $F_{\mu\nu}$ also obeys $\p_{[\mu}F_{\nu \lambda]}=0$.

We will also assume that the conserved current $j^\mu$ is contained inside the membrane $\M$ and hence the observer $\O$ who has access only to the region outside the stretched horizon does not see the current $j^\mu$. However, the observer will {\it see} a surface current $\J_\M^\mu$ on the membrane. Let us start with the action for the observer $\O$
\be\label{action1}
S_\O=-\frac{1}{4}\int_A \sqrt{g}d^4x  F_{\mu\nu}F^{\mu \nu}+\int_\M \sqrt{-h}  d^{3}x \J_{\M;\mu} A^\mu \ .
\ee
Note that the action is invariant under the gauge transformation: $A_\mu \rightarrow A_\mu +\p_\mu \alpha $ only if $\J_{\M;\mu}n^\mu=0$, where $n_\mu$ is the outgoing unit normal vector on $\M$. In order for the observer $\O$ to recover the vacuum Maxwell's equations, the boundary terms on $\M$ should cancel out and hence from equation (\ref{source}) we obtain
\be\label{Mcurrent}
 \J_\M^\mu=n_\nu F^{\mu\nu}|_\M\ .
\ee
It is obvious that $\J_{\M;\mu}n^\mu=0$ and hence action (\ref{action1}) is invariant under gauge transformations.

\subsection{B. Black hole horizon and electrical conductivity}\label{conductivity}
Again consider a fiducial observer $\O$ hovering just outside a $(3+1)-$dimensional black hole. For such an observer, the near horizon geometry is a good approximation and the metric takes the Rindler form
\begin{equation}\label{rindler}
ds^2=-\rho^2 d\omega^2+d\rho^2+dy^2+dz^2 \,.
\end{equation}
This can be regarded as a portion of Minkowski space, formally known as the Rindler wedge. In particular, under the redefinitions
\begin{eqnarray}
t= \rho \sinh \omega\ , \qquad x= \rho \cosh \omega\, ,
\end{eqnarray}
one arrives to the more familiar metric
\be
ds^2=-dt^2+dx^2+dy^2+dz^2\,.
\ee
For the observer $\O$, who follows orbits of the time-like Killing vector $\xi=\partial_\omega$, there is a horizon at the edge of the Rindler wedge, $x=|t|$, or equivalently, $\rho=0$. We will replace the mathematical horizon by the stretched horizon located at $\rho=\epsilon$, where $\epsilon$ is about one Planck length. 

Since, $4-$velocity $U^\mu$ of the observer $\O$ is singular near the horizon, the electric and magnetic fields $\mathbf{E}$ and $\mathbf{B}$ as measured by $\O$ ($e^{\nu \mu \alpha \beta}$ is the Levi-Civita tensor)
\begin{equation}\label{electric}
\mathbf{E}^\mu=F^{\mu \nu}U_\nu\ , \qquad \mathbf{B}^\mu=\frac{1}{2}e^{\nu \mu \alpha \beta}F_{\alpha \beta}U_{\nu}
\end{equation}
can be singular in general. In order to understand the behavior of $\mathbf{E}$ and $\mathbf{B}$ on the horizon let us consider a freely falling observer (FFO) $P$:
\begin{align}
& \text{FFO}\ P:\qquad x=a,\ y=z=0\ .
\end{align}
A freely falling observer does not see the coordinate singularity and hence electric and magnetic fields $\mathbf{E}_{P}$ and $\mathbf{B}_P$ as measured by $P$ should be non-singular. Relating $\mathbf{E}$ and $\mathbf{B}$ with $\mathbf{E}_{P}$ and $\mathbf{B}_P$, we obtain
\begin{align}\label{bc}
\mathbf{E}^\rho|_\M, \mathbf{B}^\rho|_\M= \O(1)\ , \nonumber\\
\mathbf{E}^y|_\M+\mathbf{B}^z|_\M=\O(\epsilon)\ , \\
\mathbf{E}^z|_\M-\mathbf{B}^y|_\M=\O(\epsilon)\ . \nonumber
\end{align}
It is important to note that the above relations can be thought of as ingoing boundary conditions for the electromagnetic radiations and it is a consequence of the fact that black holes behave as  perfect absorbers.

Therefore, from equation (\ref{Mcurrent}), we obtain,
\begin{equation}
 \J_\M^y=\mathbf{E}^y|_\M\ , \qquad  \J_\M^z=\mathbf{E}^z|_\M\ .
\end{equation}
The black hole horizon behaves like an Ohmic conductor with conductivity
\be\label{con}
\sigma=1\ .
\ee
That is, the surface resistivity of the black hole is $r=1/\sigma\approx 377 \Omega$ \cite{Price:1986yy,Thorne:1986iy}.

We will end this section with a brief discussion on Ohmic dissipation in the stretched horizon because of the electromagnetic fields on the horizon. For a Schwarzschild black hole of mass $M$, the first law of thermodynamics states
\be
TdS=dM
\ee
where $T$ is the temperature and $S$ is the entropy of the black hole. Presence of electromagnetic fields on the horizon can increase the mass and hence the entropy of the black hole \cite{Thorne:1986iy}
\be\label{dsdt}
\frac{dM}{dt}=T\frac{dS}{dt}=-\int_\M \vec{S}_H.d\vec{A}\ .
\ee
$\vec{S}_H$ is the renormalized Poynting flux
\be
\vec{S}_H=\epsilon^2 \left(\mathbf{E} \times \mathbf{B}\right)_\M\ .
\ee
$\vec{S}_H$ is the amount of red-shifted energy (as measured at infinity) crossing a unit area per unit time at infinity. Equation (\ref{bc}), leads to the famous result of the Joule heating law for black hole horizon
\be\label{jheating}
\frac{dM}{dt}=T\frac{dS}{dt}=\int_\M \epsilon^2 r \vec{\J}_{\M}^2 dA\ .
\ee
\section{III. Electrodynamics $\theta$-term and the membrane paradigm}

The gauge invariance of electrodynamics allows for a $\theta$-term in the action
\begin{align}\label{actioncs}
S=&\int \sqrt{g}d^4x \left[- \frac{1}{4}F_{\mu\nu}F^{\mu \nu}+j_\mu A^\mu \right]\nonumber\\
&~~~~~~~~~~~~~~~~~+\frac{\theta}{8} \int d^4x \epsilon^{\alpha \beta \mu \nu}F_{\alpha \beta}F_{\mu \nu}
\end{align}
where one can write 
\be
\frac{\theta}{8}\epsilon^{\alpha \beta \mu \nu}F_{\alpha \beta}F_{\mu \nu}=\frac{\theta}{4}\sqrt{g}F_{\mu \nu}*F^{\mu\nu}\ .
\ee
Our convention of the Levi-Civita tensor density $\epsilon^{\alpha \beta \mu \nu}$ is the following: $\epsilon^{0123}=1$, $\epsilon_{0123}=-g$. 

The electrodynamics $\theta$-term is a total derivative and hence it does not affect the classical equations of motion. Therefore, this term does not contribute even for perturbative quantum electrodynamics, which strongly indicates that the effects of the QED $\theta$-term, if any, are non-perturbative. However, by coupling this theory to gravity we will show that the electrodynamics $\theta$-term can affect the electrical properties of the stretched horizon.

Let us again imagine a stretched horizon $\M$ that divides the whole space-time in two regions and an observer $\O$ who has access only to the region outside the horizon. The action for the observer $\O$ is
\begin{align}
S_\O=\int_A \sqrt{g}d^4x \left[- \frac{1}{4}F_{\mu\nu}F^{\mu \nu}+\frac{\theta}{4}F_{\mu \nu}*F^{\mu\nu} \right]\nonumber\\
+\int_\M \sqrt{-h}  d^{3}x \J_{\M;\mu} A^\mu \ .
\end{align}
Therefore from equation (\ref{source}) the membrane surface current is given by,
\be\label{hcurrent}
 \J_\M^\mu=\left(n_\nu F^{\mu\nu}-\theta\ n_\nu *F^{\mu\nu}\right)|_\M\ .
\ee

\subsection{A. Hall conductivity}
Following the discussion of section(IIB), equations (\ref{bc}) and (\ref{hcurrent}), now lead to
\be
\begin{pmatrix}
 \J_\M^y \\
  \J_\M^z  
 \end{pmatrix}=
\begin{pmatrix}
  \sigma & -\theta \\
  \theta &~~ \sigma 
 \end{pmatrix}
 \begin{pmatrix}
   \mathbf{E}^y_\M\\
  \mathbf{E}^z_\M  
 \end{pmatrix}\ .
\ee
Therefore, surface Hall conductance of the black hole horizon is
\be
\sigma_{zy}=-\sigma_{yz}=\theta \approx \theta(377\Omega)^{-1}\ 
\ee
and in principle one can find out the value of the $\theta$-angle by measuring the Hall conductivity of the black hole horizon.

From equation (\ref{dsdt}), it is obvious that the presence of the $\theta$-term does not contribute to the  increase of entropy of the black hole. However, it can be easily checked that the stretched horizon now does not follow the standard Joule heating law (\ref{jheating}). Instead it obeys
\be
\frac{dM}{dt}=T\frac{dS}{dt}=\int_\M \epsilon^2 \left(\frac{\sigma}{\theta^2+\sigma^2}\right) \vec{\J}_{\M}^2 dA\ .
\ee
\section{IV. Hall scrambling of charges on the stretched horizon}
Scrambling is the process by which a localized perturbation of a system spreads out into the whole system. In quantum mechanics, information contained  inside a small subsystem of a bigger system, is said to be scrambled when the small subsystem becomes entangled with the rest of the system. And scrambling time $t_s$ is defined as the time it takes for a localized perturbation to 
become fully scrambled such that the information it contains can only be recovered by examining nearly all the degrees of freedom. In a local quantum field theory, scrambling time $t_s$ is expected to be at least as long as the diffusion time. Consequently, for a strongly correlated quantum fluid in $d$-spatial dimensions and at temperature $T$, the scrambling time satisfies 
\be\label{bound}
t_s T \ge c \hbar S^{2/d}\ ,
\ee
where $c$ is some dimensionless constant and $S$ is the total entropy. In \cite{Sekino:2008he,Susskind:2011ap}, it has been argued that this is a universal bound on the scrambling time. Hence, it is indeed remarkable that information scrambles on black hole and de Sitter horizons exponentially fast
\be
t_s T \approx \hbar \ln S
\ee
violating the bound (\ref{bound}). This unusual process is famously known as ``fast-scrambling" and it strongly indicates that the microscopic description of scrambling of information on static horizons must involve non-local degrees of freedom \cite{Sekino:2008he,Susskind:2011ap}. Non-locality is indeed essential for fast scrambling \cite{nonlocal1,nonlocal2,nonlocal3,nonlocal4,nonlocal5} and in fact it is well known that non-local interactions can increase the level of entanglement among different degrees of freedom of a theory \cite{entanglement1,entanglement2,entanglement3}.

In this section, we will focus on scrambling of point-charges on the horizon in the presence of the electrodynamics $\theta$-term. We will argue that the $\theta$-angle affects the way charges scramble on the horizon, in particular, it introduces vortices without changing the scrambling rate -- we will call this phenomenon ``Hall scrambling". Let us note that a microscopic description of fast-scrambling should be able to explain the origin of Hall-scrambling in the presence of the $\theta$-angle.

\subsection{A. Rindler coordinates}
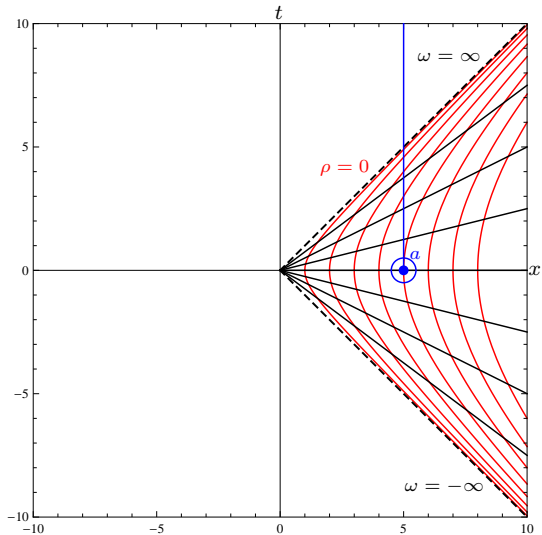
\begin{figure}[!htbp]
\begin{center}
  \includegraphics[width=7cm]{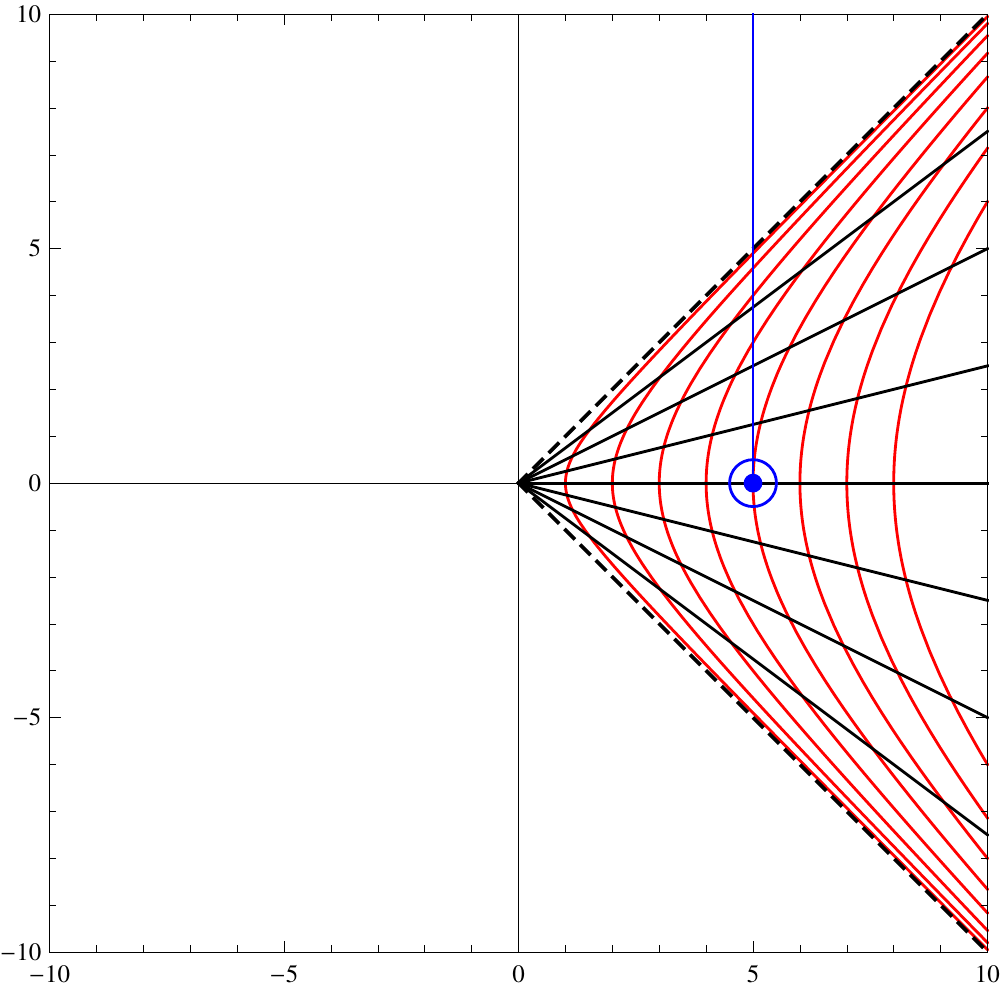}
   \begin{picture}(0,0)
 \put(-5,99){\footnotesize{$x$}}
  \put(-101,196){\footnotesize{$t$}}
  \put(-47,180){\scriptsize{$\omega=\infty$}}
  \put(-52,17){\scriptsize{$\omega=-\infty$}}
  \put(-84,138){\scriptsize{\color{red} $\rho=0$}}
  \put(-50,105){\scriptsize{\color{blue} $a$}}
\end{picture}
  \end{center}
\caption{Rindler coordinates plotted on a Minkowski diagram. The dashed lines correspond to the Rindler horizons. Constant-$\rho$ and constant-$\omega$ lines are shown in red and black, respectively. The blue line corresponds to the worldline for a free falling charge.\label{fig1}}
\end{figure}

Let us again consider an accelerated observer in Minkowski space. For such an observer, the metric takes the Rindler form (\ref{rindler}). We again replace the mathematical horizon by the stretched horizon at $\rho=\epsilon$, where $\epsilon$ is about one Planck length. We will consider a single charge which is stationary in the Minkowski frame \cite{lindesay}
\begin{align}
& \text{Charge}\ Q:\qquad x=a,\ y=z=0\ .
\end{align}
However, in the Rindler frame, the charge is falling into the horizon. Figure \ref{fig1} represents schematically this situation.

To compute the surface current, we need to compute $F^{\mu\nu}_R$ in the Rindler frame
 \be
 \J_\M^\mu= \left(F^{\mu \rho}_R-\theta *F_R^{\mu\rho}\right)|_\M\ .
\ee
At any given time the Rindler coordinates are related to the Minkowski coordinates by a boost along the $x$-axis. In Minkowski coordinates:
\begin{align}
& F_M^{01}=\frac{Q(x-a)}{4\pi \left[(x-a)^2+y^2+z^2\right]^{3/2}}=-F_M^{10}\ ,\\
& F_M^{02}=\frac{Qy}{4\pi \left[(x-a)^2+y^2+z^2\right]^{3/2}}=-F_M^{20}\ ,\\
& F_M^{03}=\frac{Qz}{4\pi \left[(x-a)^2+y^2+z^2\right]^{3/2}}=-F_M^{30}\ 
\end{align}
 and all the other components are zero. Now one can compute $F^{\mu\nu}_R$ by performing the change of coordinates. That leads to:
\begin{align}
&\J_\M^\omega=\frac{Q(\epsilon \cosh \omega-a)}{4\pi \epsilon \left[(\epsilon \cosh \omega-a)^2+y^2+z^2\right]^{3/2}}\ , \\
&\J_\M^\rho=0\ , \\
&\J_\M^y=\frac{Q\left(y \sinh \omega-\theta z \cosh \omega\right)}{4\pi  \left[(\epsilon \cosh \omega-a)^2+y^2+z^2\right]^{3/2}}\ ,\label{J2} \\
&\J_\M^z=\frac{Q\left(z \sinh \omega+\theta y \cosh\omega\right)}{4\pi  \left[(\epsilon \cosh \omega-a)^2+y^2+z^2\right]^{3/2}}\ .\label{J3}
\end{align} 
We are mainly interested in the current density on the stretched horizon after the point charge $Q$ crosses the stretched horizon. An observer hovering just outside the stretched horizon will measure a surface charge density $\rho_H (y,z)=\epsilon \J_\M^0$ on the horizon. Note that the surface charge density $\rho_H (y,z)$ does not receive any correction in the presence of the $\theta$-term.

The point charge crosses the horizon at the Rindler time $\omega=\omega_1$, where
\be
\cosh \omega_1=\frac{a}{\epsilon}.
\ee 
In the limit $\omega\rightarrow \omega_1$ (but $\omega<\omega_1$), we obtain
\be
\rho_H (y,z)=- \frac{Q}{2} \ \delta\left(y\right)\delta\left(z\right)\ .
\ee
In the limit $\omega\rightarrow \omega_1$ (but $\omega>\omega_1$)
\be
\rho_H= \frac{Q}{2} \ \delta\left(y\right)\delta\left(z\right)\ .
\ee
Note that there is a discontinuity at $\omega=\omega_1$. The total charge on the horizon is given by ($\omega>\omega_1$):
\be
Q_H=\int dydz \rho_H (y,z)=\frac{Q}{2}\ .
\ee

An observer $\O$ hovering near the the horizon will measure the current $\J_\M^\mu$ and electric field $\mathbf{E}$ (\ref{electric}):
\be
\mathbf{E}^y=(\cosh \omega) F^{02}_M|_\M, \qquad \mathbf{E}^z=(\cosh \omega) F^{03}_M|_\M\ .
\ee
Finally equations (\ref{J2},\ref{J3}) can be rewritten in the following way:
\be
\begin{pmatrix}
 \J_\M^y \\
  \J_\M^z  
 \end{pmatrix}=
\begin{pmatrix}
  \tanh \omega & -\theta \\
  \theta & \tanh \omega 
 \end{pmatrix}
 \begin{pmatrix}
   \mathbf{E}^y\\
  \mathbf{E}^z  
 \end{pmatrix}\ .
\ee

Therefore, in this case, conductivity of the stretched horizon, as measured by the observer $\O$ is time-dependent 
\be\label{con}
\sigma=\tanh \omega\approx 1-2 e^{-2\omega}\ .
\ee

\subsubsection{\bf Late time behavior: Scrambling}
In the limit $\omega>> \omega_1$ we obtain:
\begin{align}
&\rho_H (y,z)=\frac{\epsilon Q   e^{-2\omega}}{\pi  \left[\epsilon^2+r_\perp^2\right]^{3/2}}\ , \label{r1}\\
&\J_\M^y=\frac{(y-\theta z) Q   e^{-2\omega}}{\pi  \left[\epsilon^2+r_\perp^2\right]^{3/2}}\ ,\label{r2} \\
&\J_\M^z=\frac{(z+\theta y) Q   e^{-2\omega}}{\pi  \left[\epsilon^2+r_\perp^2\right]^{3/2}}\ ,\label{r3}
\end{align} 
where $r_\perp^2= 4 e^{-2\omega}(y^2+z^2)$. And in this limit the conductivities of the stretched horizon are constants: $\sigma_{yy}=\sigma_{zz}\approx 1$, $\sigma_{zy}=-\sigma_{yz}=\theta$ and the horizon behaves like an ideal Hall conductor.

Before proceeding further, a few comments are in order. From equation (\ref{r1}), it is clear that the presence of the $\theta$-term does not change the scrambling time. On the other hand, equations (\ref{r2}) and (\ref{r3}) indicate that the $\theta$-angle affects the way the charge scramble on the horizon. Let us first represent $\{y,z\}$-plane in the polar coordinates: $\{b,\phi\}$, where $y=b \cos \phi$ and $z=b\sin \phi$, as usual. When $\theta=0$, $\phi$-component of the horizon current $\J^\phi_\M=0$. Whereas, for $\theta \neq 0$, $\J^\phi_\M=\theta \J^b_\M$ and $\vec{\nabla}\times\vec{\J}_\M\neq 0$ which clearly indicate the presence of vortices; this is a direct consequence of the non-zero Hall conductivity. 

\subsection{B. Schwarzschild black hole}

The metric of a Schwarzschild black hole in Kruskal-Szekeres coordinates is 
\begin{align}
ds^2=\frac{32G^3m^3}{r}e^{-\frac{r}{2Gm}}\left(-dV^2+dU^2\right)~~~~~~~&\nonumber\\
  +r^2 \left(d\bar{\theta}^2 + \sin^2\bar{\theta} d\phi^2\right)&\ ,
\end{align}
where $r$ is the radial Schwarzschild coordinate
\begin{equation}
V^2-U^2=\left(1-\frac{r}{2Gm}\right)e^{\frac{r}{2Gm}}
\end{equation}
and we are using the symbol $\bar{\theta}$ for angular coordinate to differentiate it from the $\theta$-angle of the action (\ref{actioncs}). The event horizon is located at $V=\pm U$ and the curvature singularity is at $V^2-U^2=1$.
\begin{figure}[!htbp]
\begin{center}
  \includegraphics[width=8cm]{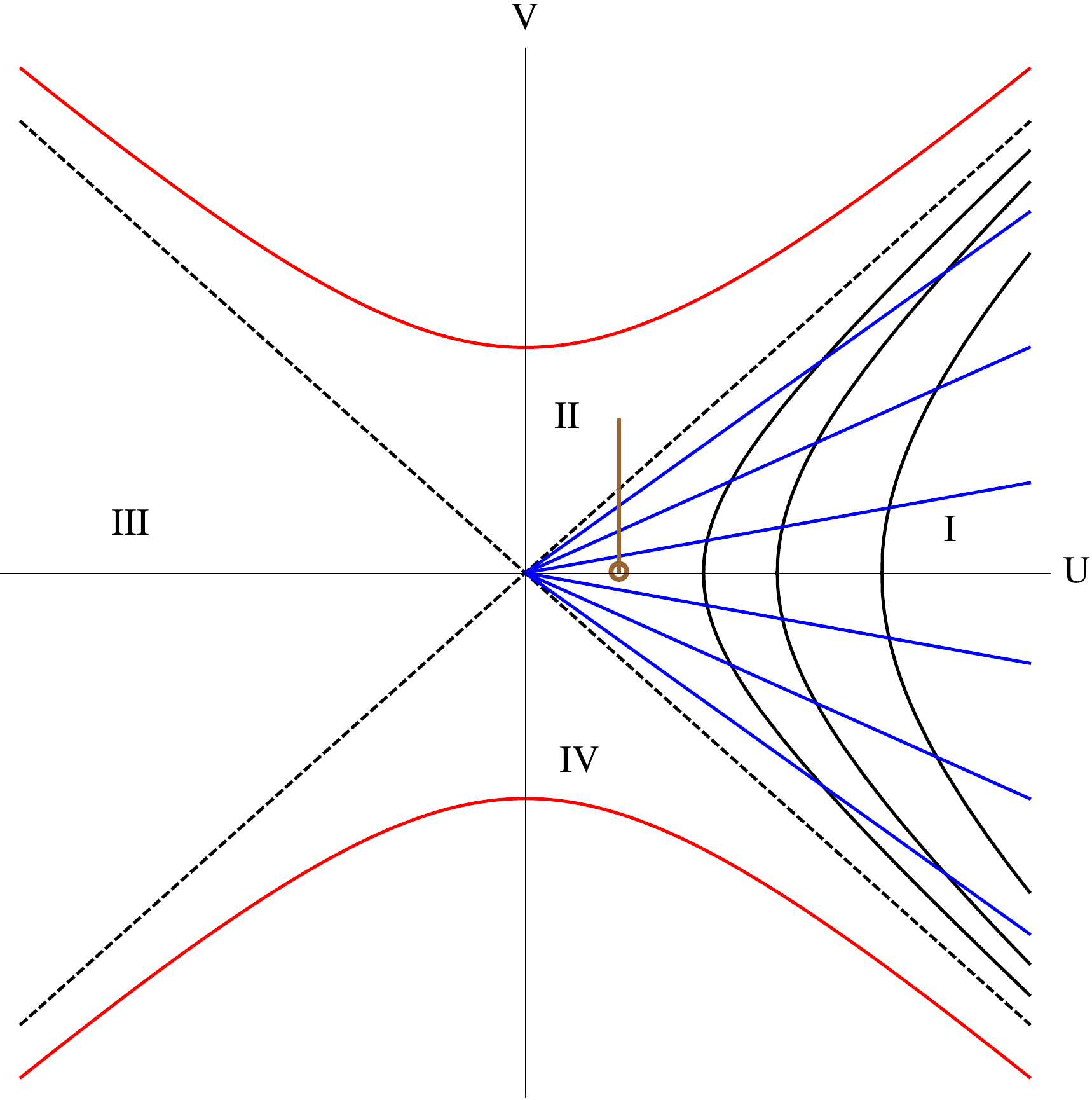}
  \end{center}
\caption{Maximal analytic extension of Schwarzschild solution in Kruskal-Szekeres coordinates. Region I is the Schwarzschild patch, where constant $r$-lines are shown in black and constant $t$-lines are shown in Blue. The solid red lines represent the singularity. The world line of a free falling charge from $V=0$ is shown in brown.}
\end{figure}
In Schwarzschild coordinates, the metric is 
\begin{align}
ds^2=-\left(1-\frac{r}{2Gm}\right)&dt^2+\frac{dr^2}{\left(1-\frac{r}{2Gm}\right)}\nonumber\\
&+r^2 \left(d\bar{\theta}^2 + \sin^2\bar{\theta} d\phi^2\right)\ .
\end{align}
For $r>2Gm$, coordinates $\{U,V\}$ and $\{r,T\}$ are related in the following way
\begin{align}
V=& \left(\frac{r}{2Gm}-1\right)^{1/2}e^{\frac{r}{4Gm}}\sinh \left(\frac{t}{4Gm}\right)\ , \\
U=& \left(\frac{r}{2Gm}-1\right)^{1/2}e^{\frac{r}{4Gm}}\cosh \left(\frac{t}{4Gm}\right)\ .
\end{align}
Following the standard procerdure, we will replace the mathematical horizon by the stretched horizon at $r=2Gm+\delta$, where $\delta<<2Gm$.

\begin{figure}[h]
\begin{center}
\unitlength = 1mm

\subfigure[ ]{
\includegraphics[width=43mm]{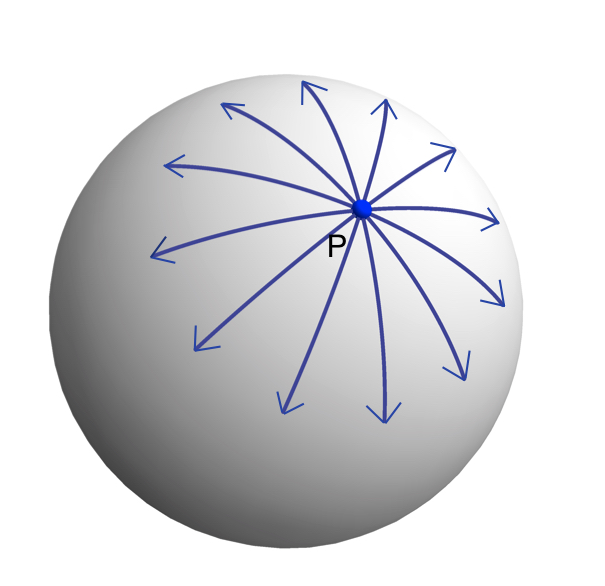}
}
\subfigure[ ]{
\includegraphics[width=55mm]{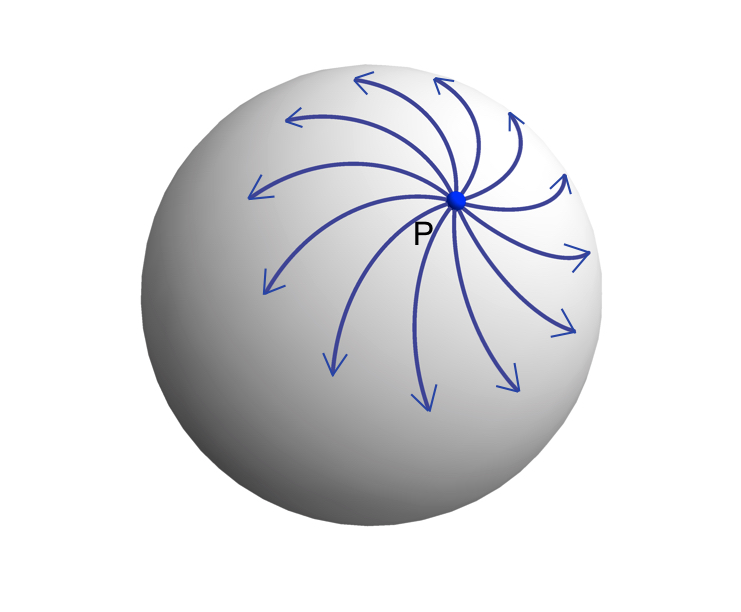}
 }
\subfigure[ ]{
\includegraphics[width=50mm]{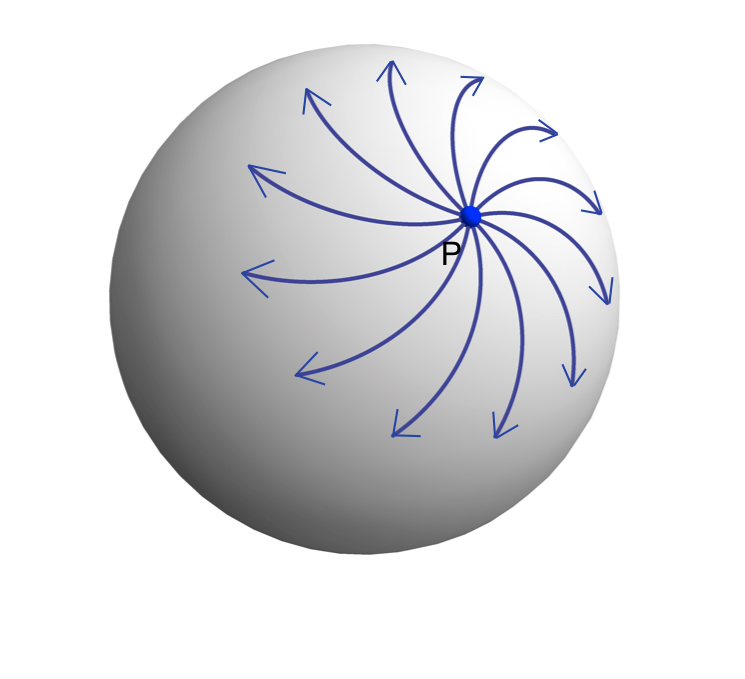}
 }
\caption{\small A schematic diagram of Hall scrambling on a black hole horizon.  Lines represent surface electric current on the black hole horizon after a positive charge is dropped at point $P$, as observed by an outside observer for (a) $\theta=0$,  (b) $\theta>0$, (c) $\theta<0$.} \label{hsfig}
\end{center}
\end{figure}

We will consider a large black hole and restrict the charges to be near the horizon, i.e. $|r/2Gm-1|<<1$, in a small angular region arbitrarily centered at $\bar{\theta}=0$. In that case, we can replace the angular part of both Kruskal-Szekeres and Schwarzschild coordinates by Cartesian coordinates
\begin{equation}
r^2 \left(d\bar{\theta}^2 + \sin^2\bar{\theta} d\phi^2\right)\approx dy^2+dz^2\ ,
\end{equation}
where,
\begin{align}
y=2mG \bar{\theta} \cos\phi\ ,\qquad z=2mG \bar{\theta} \sin\phi\ .
\end{align}
We will now define two sets of coordinates to describe the near horizon region of the Schwarzschild black hole:
\begin{align}
&\rho=4Gm\sqrt{1-\frac{2Gm}{r}}, \qquad \omega=\frac{t}{4Gm}\ .\\
&X= 4Gme U, \qquad T= 4Gme V\ .
\end{align}
Where $e$ is the Euler's number. Therefore, in terms of $\{T,X,y,z\}$, the Kruskal-Szekeres metric becomes (near the horizon)
\begin{equation}
ds^2\approx-dT^2+dX^2+dy^2+dz^2\ .
\end{equation}
Similarly, the Schwarzschild metric leads to
\begin{equation}
ds^2\approx-\rho^2d\omega^2+d\rho^2+dy^2+dz^2\ .
\end{equation}
Therefore, we can use the results of the Rindler case to analyze smearing of charges on the Schwarzschild horizon. Again we have a charge $Q$ and initial conditions are given at $V=0$:
\begin{align}
Q:& \qquad U=U_0,\qquad \bar{\theta}=0\ .
\end{align}
In the near horizon approximation, we can map this problem to the problem of scrambling in Rindler spacetime with 
\begin{align}
a=4GmeU_0,  \qquad \epsilon=2\sqrt{2Gm\delta}\ .
\end{align}
Therefore, the Schwarzschild observer $\O$ can see the charge $Q$ for $t<t_1$. At Schwarzschild time $t=t_1$, the Schwarzschild observer will see that the charge density on the stretched horizon is localized at a point , where
\be
\cosh\left(\frac{t_1}{4Gm}\right)=\frac{e U_0 \sqrt{2Gm}}{\sqrt{\delta}}\ .
\ee
Finally using (\ref{r1}-\ref{r3}), in the late time limit $t>> t_1$, (in the small angle approximation) we obtain:
\begin{align}
&\rho_H (\bar{\theta},\phi,t)=\frac{2\sqrt{2Gm\delta} Q   e^{-t/2Gm}}{\pi  \left[8G m \delta+r_\perp^2\right]^{3/2}}\ , \label{l1}\\
&\J_\M^{\bar{\theta}}(\bar{\theta},\phi,t)=\frac{2m G \bar{\theta} Q   e^{-t/2Gm}}{\pi  \left[8G m \delta+r_\perp^2\right]^{3/2}}\ , \\
&\J_\M^\phi(\bar{\theta},\phi,t)=\theta \J_\M^{\bar{\theta}}(\bar{\theta},\phi,t)\ ,
\end{align} 
where $r_\perp= 4m G \bar{\theta} e^{-t/4mG}$. So, an observer hovering outside the horizon will see Hall scrambling on the Schwarzschild horizon as expected (see figure \ref{hsfig}). Note that the Hall scrambling depends only on the Rindler-like character of the horizon but not on the details of the Schwarzschild metric. We can define the scrambling time $t_s$ in the following way: it is the time at which the charge density $\rho_H (\bar{\theta},\phi,t_s)\approx Q/4\pi r_H^2$, where $r_H=2Gm$ is the radius of the horizon. Applying that for $\bar{\theta}=0$, we obtain,
\be
t_s\approx 2Gm \ln \left(\frac{2 G m}{\delta}\right)\approx 4Gm \ln \left(\frac{m}{2\pi M_P}\right)\ ,
\ee
where we have used $\epsilon\sim$ Planck length.  So the Hall scrambling does not affect the scrambling rate.

It is important to note that there is a crucial difference between the discussions of Rindler spacetime and the Schwarzschild black hole. In a Schwarzschild black hole, all freely falling objects will hit the singularity at $r=0$ in finite  Kruskal-Szekeres time. One can argue \cite{Banks:2014xja} that  when a single charge hits the singularity, the spherical symmetry will be restored and the total charge will be uniformly distributed over the stretched horizon. In the scrambling time, $t_s$ an order one perturbation will decay to size $ \sim M_P/m$, and all trace of it will be lost \cite{Sekino:2008he, lindesay}. Hence, this should be the time scale for any classical fields on the horizon to become spherically symmetric.

\subsection{C. Cosmological horizon: Hall scrambling and de-scrambling}
We can easily extend the discussion of this section for cosmological horizons. Before, we use the results of the Rindler calculations, we should modify equation (\ref{source}) for cosmological horizons. We can simply follow the discussions of section (II) and conclude that
\be
\J_\M^I= -\left[n_\mu \frac{\partial \L}{\partial \left(\del_\mu \phi_I \right)}\right]_\M
\ee
where, $n_\mu$ is the outward pointing normal vector to the stretched horizon $\M$. Without the $\theta$-angle, equation (\ref{Mcurrent}) now becomes
\be\label{dscurrent}
 \J_\M^\mu=-n_\nu F^{\mu\nu}|_\M\ .
\ee

We will mainly focus on de Sitter space. It was pointed out in \cite{Susskind:2011ap} that charges also fast scramble across the de Sitter horizon. It is not very surprising since the near horizon region of de Sitter space is Rindler-like. However, it strongly suggests that the holographic description of a causal patch of de Sitter space must involves non-local degrees of freedom \cite{Susskind:2011ap}. A comoving observer $\O$ in de Sitter space only sees a static patch of de Sitter space with metric,
\begin{align}\label{ds}
ds^2= -(1-H^2 r^2)&dt^2+\frac{dr^2}{1-H^2 r^2}\nonumber\\
&+r^2 \left(d\bar{\theta}^2 + \sin^2\bar{\theta} d\phi^2\right)\ ,
\end{align}
where, cosmological constant is given by $\Lambda=3H^2$. Rest is straightforward: we will replace the mathematical horizon by the stretched horizon at $rH=1-\delta$. Let us now imagine a charge $Q$ that the observer $\O$ can see  for $t<t_1$ and from (\ref{dscurrent}) it is obvious that the total induced charge on the horizon $Q_H=-Q$. At time $t=t_1$, the charge $Q$ hits the stretched horizon at a point $\bar{\theta}=0$. For $t>t_1$, unlike the Schwarzschild case, the Gauss's law tells us: total charge on the de Sitter horizon $Q_H=0$. However, we will not be able to see that from the Rindler approximation which is valid only for small angle limit. Note that, following \cite{Carney:2013wpa} it is possible to perform an exact calculation in de Sitter space  but we will not attempt it in this paper. 

In the near horizon limit, at small angle approximation the metric (\ref{ds}) becomes Rindler-like with
\begin{align}
&\rho=\cos^{-1} (Hr)\ , \qquad \omega=Ht\\ \nonumber
&y=H^{-1} \bar{\theta} \cos\phi\ ,\qquad z=H^{-1} \bar{\theta} \sin\phi\ 
\end{align}
and $\epsilon=\sqrt{\delta}/H$. Hence, we will see Hall scrambling on de Sitter horizon with the electrodynamics $\theta$-term. In particular, in the late time limit $t>> t_1$, we obtain:
\begin{align}
&\rho_H (\bar{\theta},\phi,t)=-\frac{\sqrt{\delta} Q   e^{-2H t}}{\pi  H \left[\delta/H^2+r_\perp^2\right]^{3/2}}\ , \label{d1}\\
&\J_\M^{\bar{\theta}}(\bar{\theta},\phi,t)=-\frac{ \bar{\theta} Q   e^{-2Ht}}{\pi H \left[\delta/H^2+r_\perp^2\right]^{3/2}}\ , \label{d2}\\
&\J_\M^\phi(\bar{\theta},\phi,t)=\theta \J_\M^{\bar{\theta}}(\bar{\theta},\phi,t)\ ,
\end{align} 
where $H r_\perp= 2 \bar{\theta} e^{-H t}$. The corresponding scrambling time is given by,
\be
t_s \approx \frac{1}{H}\ln \left(\frac{2 M_P}{H}\right)\ .
\ee
The fact that the de Sitter horizon is a fast scrambler indicates that a dual description, if exists, must be non-local in nature \cite{Susskind:2011ap} and it should also be able to provide a microscopic description of the Hall scrambling. 

Discussion of this section can easily be generalized for arbitrary cosmological horizons and one can show that in the presence of the electrodynamics $\theta$-angle point charges will Hall scramble on the apparent horizon of a co-moving observer if the expansion of the universe is accelerating. On the other hand, when the expansion of the universe is decelerating, the observer sees the charges ``{\it Hall de-scramble}" as they re-enter the horizon. However, there is a crucial difference: for arbitrary non-de Sitter cosmological expansion both scrambling and de-scrambling occur at a power law rate \cite{Carney:2013wpa}. This perhaps indicates that it may be possible to describe the process of Hall (de)-scrambling on arbitrary non-de Sitter cosmological horizons in terms of locally interacting degrees of freedom \cite{Sekino:2008he,Susskind:2011ap}.

\section{V. A$\text{d}$S/CFT}

The gauge-gravity duality or the AdS/CFT correspondence \cite{Maldacena:1997re, Witten:1998qj, Gubser:1998bc, Aharony:1999ti} has been successful at providing us with theoretical control over a large class of gauge theories. It is a remarkable achievement to compute observables of strongly coupled large-N gauge theories in $d$-dimensions by performing some classical gravity computations in $(d+1)$-dimensions. At finite temperature, gravity duals of these field theories have black holes with horizons and at very long length scales the most dominant contributions come from the near horizon region \cite{Fischler:2012ca}. So, it is somewhat expected that there is some connection between the low energy hydrodynamic description of these strongly coupled theories and the membrane paradigm fluid on the horizon \cite{Kovtun:2003wp,Son:2007vk}. This connection was made precise in \cite{Iqbal:2008by}, where the authors have shown that the low frequency limit of linear response of the fluid of the boundary theory is given by the that of the membrane paradigm fluid on the black hole horizon.

In particular, let us consider U(1) gauge field in AdS-Schwarzschild in $(3+1)-$dimensions with the action
\begin{align}
S=\int d^4x \left[- \frac{\sqrt{g}}{4g_{3+1}^2}F_{\mu\nu}F^{\mu \nu} +\frac{\theta}{8}  \epsilon^{\alpha \beta \mu \nu}F_{\alpha \beta}F_{\mu \nu}\right]\ .
\end{align}
The U(1) gauge field in the bulk is dual to a conserved current $j^i$ in the boundary theory and DC conductivities are given by
\be
\sigma^{ij}=-\lim_{\omega, \mathbf{k} \rightarrow 0} \frac{G_R^{ij}(\mathbf{k})}{i \omega}\ ,
\ee
where, $G_R^{ij}(\mathbf{k})$ is the retarded Green function of boundary current $j^i$. As shown in \cite{Iqbal:2008by}, the DC conductivity $\sigma^{ij}$ of the strongly coupled $(2+1)-$dimensional dual theory is given by that of the membrane paradigm fluid on the horizon of the Schwarzschild black hole in AdS. Since the near horizon region of AdS-Schwarzschild is Rindler-like, we have
\be
\sigma^{11}=\sigma^{22}=\frac{1}{g_{3+1}^2}\ , \qquad \sigma^{21}=-\sigma^{12}=\theta\ .
\ee

Let us end by briefly discussing scrambling in the context of the AdS/CFT correspondence. Consider the prototype case of $\N=4$ super-Yang-Mills with gauge group $SU(N)$ on a unit 3-sphere at finite temperature. The AdS/CFT correspondence relates this theory to type IIB string theory on asymptotically AdS$_5\times S^5$ spacetime. In the limit $N\gg1, \lambda=g_{YM}^2 N\gg1$, the theory can be well approximated by the classical supergravity with a large AdS black hole. Since the boundary field theory is local, it is obvious that the boundary theory is not a fast scrambler. However, as argued in \cite{Sekino:2008he}, on length scales smaller than the AdS radius $R_{AdS}$, AdS space is a fast scrambler. Any localized perturbation on length scales smaller than $R_{AdS}$ in the gravity side is equivalent to a localized perturbation of a very small subset of $N^2$ degrees of freedom of the dual CFT. So, any perturbation of a very small subset of $N^2$ degrees of freedom at some point first will fast scramble among $N^2$ degrees of freedom at that point. Then it will scramble (not so fast) over the entire sphere and hence the total scrambling time is  \cite{Sekino:2008he}
\be
t_s\approx c_0 + \frac{c}{T} \log N\ ,
\ee 
where, $c_0$ is some $\O(1)$ number and $c$ is some dimensionless numerical constant. As we have discussed before, when the dual gravity theory is in $(3+1)$-dimensions, we can have a non-zero $\theta$-angle. It will be interesting to understand Hall scrambling in the context of the gauge-gravity duality.
\section{VI. Conclusions}
We have shown that in principle one can find out the value of the electrodynamics $\theta$-angle by dropping charged particles into a black hole and observing how the perturbation scrambles on the horizon. This strongly suggests that any sensible quantum theory of the black hole needs to be able to provide a microscopic description of Hall-scrambling in the presence of the $\theta$-angle. It will be extremely interesting to see if this effect has any astrophysical consequence. 

Experiments set a rather strong limit on QCD $\theta$-angle $|\theta_{QCD}|<<10^{-9}$, however, at low energies, the theta angles for QED and QCD appear as independent parameters and hence there is no reason to expect QED-$\theta \sim \theta_{QCD}$. In grand unified theories such as for example SU(5), the $\theta$-angles for QED and QCD are related by the renormalization group. However, even in the context of grand unification,  in the presence of a mechanism involving axions to solve the strong CP-problem,  it remains a challenge to relate the $\theta$-angles for QED and QCD at energy scales far below the GUT scale.

Our conclusions depend only on the Rindler-like character of the horizon and hence they are valid for any horizon which is locally Rindler-like. However, it is indeed puzzling that an in-falling observer will not see any effects of the electrodynamics $\theta$-angle whatsoever. 

\section*{Acknowledgments}
We would like to thank Tom Banks and Daniel Harlow for the helpful conversations. SK would like to acknowledge Marco Farina, Tom Hartman, Bithika Jain and Sachin Jain for the useful discussions. This material is based upon work supported by the National Science Foundation under Grant Numbers PHY-1316033 and by the Texas Cosmology Center, which is supported by the College of Natural Sciences and the Department of Astronomy at the University of Texas at Austin and the McDonald Observatory. The work of SK was also supported by NSF grant PHY-1316222.

\end{document}